\documentclass[
10pt,
twocolumn,
amsmath,
amssymb,
aip,
cha,
]{revtex4-2}
\usepackage{graphicx} 
\usepackage{dcolumn} 
\usepackage[T1]{fontenc}
\usepackage{mathptmx}
\usepackage{float}
\usepackage{xcolor}
\usepackage{mathtools}

\begin{document}
\title{Adjoint-based reduction approach for position and phase control of chimera states}

\author{Petar Mircheski}
\email{mircheski.p.d1e6@m.isct.ac.jp}

\affiliation{Department of Systems and Control Engineering, Institute of Science Tokyo, Tokyo 152-8552, Japan}

\author{Hiroya Nakao}

\affiliation{Department of Systems and Control Engineering, Institute of Science Tokyo, Tokyo 152-8552, Japan}

\affiliation{Research Center for Autonomous Systems Materialogy, Institute of Science Tokyo, Kanagawa 226-8501, Japan}

\date{\today}

\begin{abstract}
Chimera states, in which coherent and incoherent oscillator groups coexist within the same system of coupled oscillators, represent a hallmark of nonlinear collective dynamics.
In this work, we develop a reduced-order description of the position and phase of a stationary chimera in a nonlocally coupled ring of Kuramoto-Sakaguchi oscillators.
The chimera state corresponds to a stationary solution of a nonlocal field equation for the complex order parameter.
This equation possesses a two-parameter family of neutrally stable solutions associated with continuous symmetries under spatial translation on the ring and global phase shifts of the oscillators.
By introducing position and phase sensitivity functions, obtained as zero eigenfunctions of the adjoint linearized field equation, the spatiotemporal dynamics of the chimera state are reduced to a pair of ordinary differential equations for its position and phase.
The reduced equations predict how a weak applied field drives the chimera state, providing a method to identify the optimal field profiles for control.
Through numerical simulations, we validate the predicted response of the chimera to spatial perturbations and demonstrate position control of chimera via frequency modulations of the oscillators.
\end{abstract}

\maketitle
\section{Introduction}

Chimera states, anomalous spatiotemporal patterns characterized by coexisting coherent and incoherent domains within a single oscillator population, have attracted considerable attention as a counterintuitive manifestation of symmetry breaking in coupled oscillator systems~\cite{Kuramoto1995,Kuramoto1997,Kuramoto1998,nakao1999anomalous,kuramoto2002coexistence,abrams2004chimera,omel2010chimera, abrams2004chimera, laing2009chimera0, hizanidis2014chimera, maistrenko2015chimera, xie2015chimera, chabanol1997collective, omel2012stationary, omelchenko2013coherence, yao2019self, zur2018chimera, xie2014multicluster, kundu2022higher, muolo2024phase}.
Since the first analysis of stationary chimera states in nonlocally coupled phase oscillators by Kuramoto and Battogtokh~\cite{kuramoto2002coexistence,kuramoto2006mean}, chimera states have since been observed across a variety of coupling schemes and oscillator types in both identical and non-identical settings~\cite{omel2010chimera, abrams2004chimera, laing2009chimera0, hizanidis2014chimera, maistrenko2015chimera, xie2015chimera, chabanol1997collective, omel2012stationary, omelchenko2013coherence, yao2019self, zur2018chimera, xie2014multicluster, kundu2022higher, muolo2024phase}.
Experimentally, chimera states have been observed in various types of systems, such as chemical oscillators~\cite{tinsley2012chimera, wickramasinghe2013spatially}, mechanical oscillators~\cite{martens2013chimera, kapitaniak2014imperfect}, electronic circuits~\cite{gambuzza2014experimental}, and optical oscillators~\cite{hagerstrom2012experimental}.
Beyond their theoretical significance, the relevance of chimera states has been discussed, e.g., in neuroscience~\cite{andrzejak2016all, bansal2019cognitive}, where several neurological disorders have been linked to excessive pathological synchrony.
\par

A wide range of methods has been proposed to regulate the onset and position of chimera states, including gradient-based control~\cite{bick2015controlling}, feedback control~\cite{omelchenko2016tweezers, sieber2014controlling}, time-delay control~\cite{sawicki2018delay, ghosh2018engineering}, pacemakers~\cite{ruzzene2019controlling}, frequency modulation~\cite{mircheski2025spatial}, and pinning control by driving a small set of oscillators~\cite{gambuzza2016pinning, isele2016controlling, muolo2025pinning}, among others.
For comprehensive reviews and further citations of recent work, see Refs.~\cite{omel2019chimerapedia, parastesh2021chimeras, zakharova2020chimera, omel2018mathematics}.
Despite this progress, most existing control approaches treat the chimera phenomenologically, designing external forcing or feedback signals based on the observed response of the chimera state rather than deriving an explicit reduced-order model of the chimera's intrinsic dynamics.
\par

In this paper, we develop an adjoint-based reduction framework for a stationary chimera state in a nonlocally coupled ring of Kuramoto-Sakaguchi oscillators~\cite{abrams2004chimera}.
The stationary chimera state corresponds to a solution of the coarse-grained nonlocal field equation derived via the Ott-Antonsen ansatz~\cite{ott2008low}.
Reflecting the underlying spatial translations on the ring and global phase shifts of the oscillators, this field equation possesses a two-parameter family of neutrally stable equivalent stationary states rather than an isolated fixed point.

We show that a pair of reduced equations for the position and phase characterizing this family of stationary states can be derived by projecting the system dynamics onto the two neutral directions.
We introduce position and phase sensitivity functions, obtained as the zero eigenfunctions of the adjoint linearized system, and validate them by direct numerical simulations of both the nonlocal field equation and the original coupled Kuramoto-Sakaguchi oscillators.
We then use the reduced equations to design a control input to steer the position of the chimera state to a prescribed target.

\section{Model Equation}

\subsection{Nonlocally coupled phase oscillators on a ring}

We investigate a spatially extended system of continuously distributed phase oscillators on a ring of length $2\pi$, governed by the following integro-differential equation:
\begin{align}
	\dot{\theta}(x, t) = \omega(x) + \int_{-\pi}^{\pi} G(x-x') \sin(\theta(x', t) - \theta(x, t) - \alpha) dx',
	\label{eq:model}
\end{align}
where the overdot $(\dot{})$ denotes the partial derivative ${\partial} / {\partial t}$ (the ordinary derivative $d/dt$, for functions depending on $t$ alone).
Here, $\theta(x,t)$ is the phase of the oscillator at position $-\pi \leq x \leq \pi$ and time $t$, and $\omega(x)$ is the natural frequency of the oscillator at $x$.
We assume periodic boundary conditions at $x=\pm \pi$ so that system~\eqref{eq:model} describes a ring of coupled oscillators.
We also assume that $\omega(x)$ is randomly and independently chosen from a probability density function $g(\omega)$ at each $x$.
The integral term represents a spatially nonlocal Kuramoto-Sakaguchi coupling specified by a kernel function $G$, with phase lag $\alpha \in (-\pi/2, \pi/2)$.
We adopt the cosine-shaped coupling kernel introduced by Abrams and Strogatz~\cite{abrams2004chimera},
\begin{align}
	G(x-x') = \frac{1}{2\pi}(1 + A \cos(x - x')), \qquad A > 0,
	\label{eq:kernel}
\end{align}
where $A$ controls the coupling strength.
By introducing the local complex order parameter as
\begin{align}
	W(x, t) = \int_{-\pi}^{\pi} G(x-x') e^{-i \theta(x', t)} dx',
\end{align}
where a nonlocal average of $e^{-i\theta}$ is used rather than the standard $e^{i\theta}$ for later convenience~\cite{laing2009chimera},
Eq.~\eqref{eq:model} can also be represented as
\begin{align}
	\dot{\theta}(x, t) & = \omega(x) - \mbox{Re} \left[ e^{i (\theta(x, t) - \beta)} \int_{-\pi}^{\pi} G(x-x') e^{- i \theta(x', t)} dx' \right]
	\cr
	                   & =  \omega(x) - \frac{1}{2} \left[ W(x, t) e^{i (\theta(x, t) - \beta )} + c.c. \right],
	\label{eq:model12}
\end{align}
where we defined $\beta = \frac{\pi}{2} - \alpha$.
\par
\begin{figure}[t]
	\centering
	\includegraphics[width=\columnwidth]{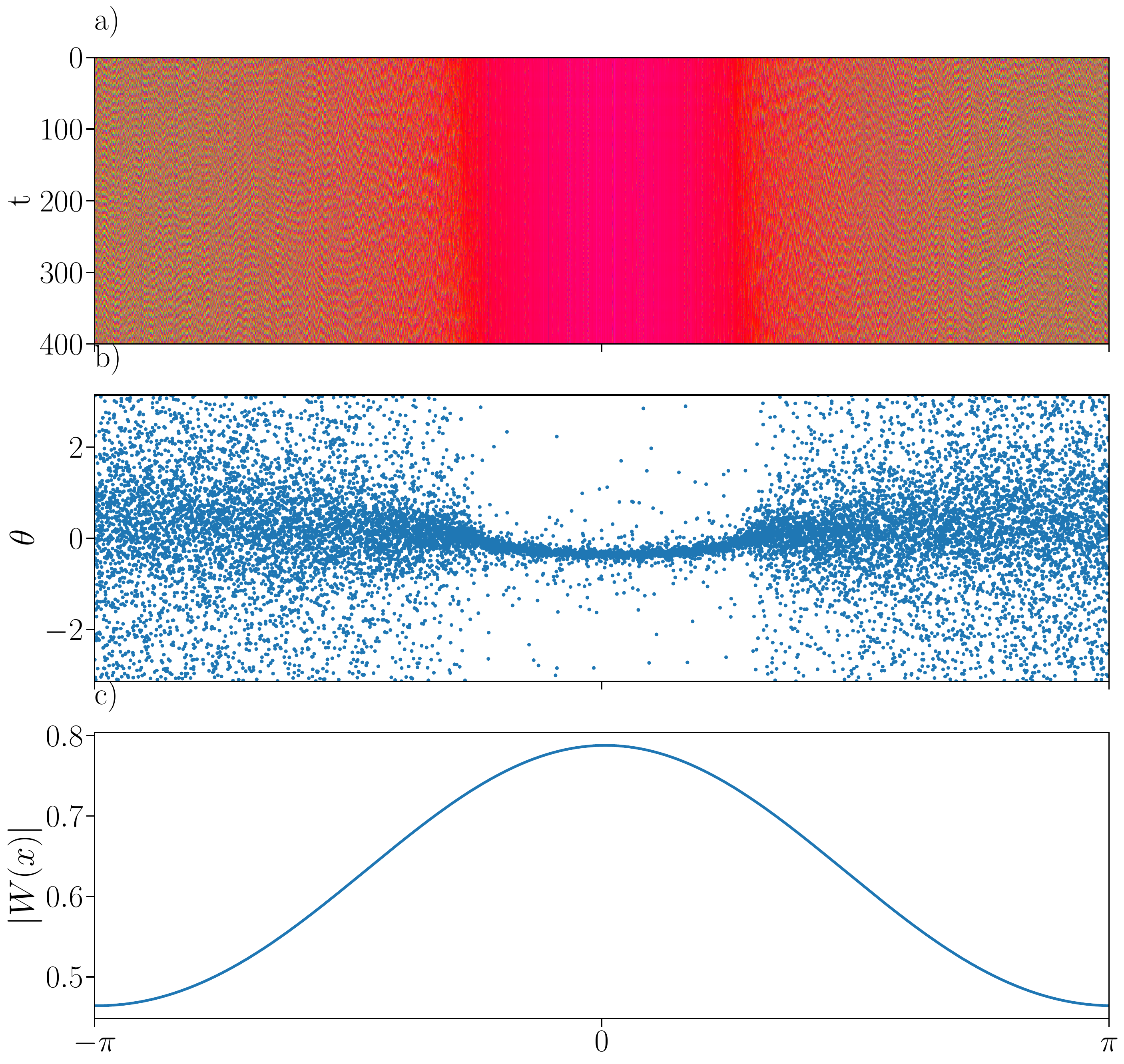}
	\caption{
		(a) Space-time plot of the phase $\theta(x,t) \in [-\pi, \pi]$ in the chimera state, obtained by simulating Eq.~\eqref{eq:model} for $N = 50\,000$ oscillators coupled through the cosine kernel, Eq.~\eqref{eq:kernel}, with $A = 0.995$ and $\alpha = \pi/2 - \beta$, $\beta = 0.16$, shown for $0 \le t \le 400$ after a transient.
		The natural frequencies $\omega(x)$ are drawn from a Lorentzian distribution centered at $\omega=0$ with half-width $D = 0.004$.
		(b) Snapshot of the phases at $t = 400$, measured relative to the global mean phase.
		(c) Time-averaged modulus of the local complex order parameter $|W(x)|$.
	}\label{fig:chimera-plot}
\end{figure}

Under the above setting with appropriate parameter values, this system is known to robustly exhibit chimera states, consisting of an incoherent domain (which we call the {\it chimera core}) of desynchronized oscillators and a coherent, synchronized domain of oscillators, as shown in Fig.~\ref{fig:chimera-plot}~\cite{abrams2004chimera}.

\subsection{Ott-Antonsen reduction to a nonlocal field equation}

As shown by Laing~\cite{laing2009chimera}, the system of continuously distributed heterogeneous oscillators, Eq.~\eqref{eq:model}, can be coarse-grained to a nonlocal field equation for the complex order parameter via the Ott-Antonsen ansatz~\cite{ott2008low}.
Here, we briefly review Laing's theory for later use.
We consider the oscillators in each infinitesimally small interval $[x, x+dx]$ as a single population.
We denote by $f(x, \omega, \theta, t) dx d\theta d\omega$ the probability that an oscillator in this interval has phase in $[\theta, \theta + d \theta]$ and frequency in $[\omega, \omega + d \omega]$.
The probability density function $f(x, \omega, \theta, t)$ is normalized as $\int_{-\infty}^{\infty} \int_{-\pi}^{\pi} f(x, \omega, \theta, t) d\omega d\theta = 1$ and gives the frequency PDF $g(\omega) = \int_{-\pi}^{\pi} f(x, \omega, \theta, t) d\theta$ when marginalized with respect to $\theta$ irrespective of $x$ and $t$.
\par
This PDF $f(x, \omega, \theta, t)$ satisfies the continuity equation
\begin{align}
	\frac{\partial f}{\partial t} + \frac{\partial}{\partial \theta}(f v) = 0,
	\label{eq:continuity}
\end{align}
where the phase velocity $v(x, \omega, \theta, t)$ of the oscillator with frequency $\omega$ is given, from Eq.~\eqref{eq:model12}, by
\begin{align}
	v = \omega - \frac{1}{2}\left[ W e^{i(\theta - \beta)} + \overline{W} e^{-i(\theta - \beta)} \right].
	\label{eq:v}
\end{align}
Here, the overbar denotes complex conjugation, and the local complex order parameter is expressed as (see Appendix A)
\begin{align}
	W(x, t)
	 & = \int_{-\pi}^{\pi} G(x-x') \int_{-\pi}^{\pi} \int_{-\infty}^{\infty} f(x', \omega', \theta', t) e^{-i \theta'} d\omega' d\theta' dx'.
	\label{eq:v2}
\end{align}
\par
The Ott-Antonsen ansatz~\cite{ott2008low} assumes that the PDF can be written in the form
\begin{align}
	f(x, \omega, \theta, t) = \frac{g(\omega)}{2\pi} \left[ 1 + \left\{ \sum_{n = 1}^{\infty} a(x, \omega, t)^n e^{in\theta} + c.c. \right\} \right],
	\label{eq:ansatz}
\end{align}
where the $n$th Fourier coefficient is given by the $n$th power of the first Fourier coefficient $a(x, \omega, t)$ (``$c.c.$'' denotes the complex conjugate of the preceding term).
Substituting Eq.~\eqref{eq:ansatz} into Eqs.~\eqref{eq:continuity} and \eqref{eq:v2}, we obtain the evolution equation 
for the first Fourier mode $a(x, \omega, t)$ as
\begin{align}
	\frac{\partial a}{\partial t} = -i\omega a + \frac{i}{2}\left[ W e^{-i\beta} + \overline{W} e^{i\beta} a^2 \right],
	\label{eq:dadt}
\end{align}
where the complex order parameter can be represented as
\begin{align}
	W(x, t)
	 & = \int_{-\pi}^{\pi} G(x - x') \int_{-\infty}^{\infty} g(\omega') a(x', \omega', t)  d\omega'  dx'\
	\cr
	 & = \left( G * \int_{-\infty}^{\infty} g(\omega') a(x', \omega', t)  d\omega' \right)(x, t),
	\label{eq:W-a}
\end{align}
In the second expression, $*$ denotes the spatial convolution with $G$ on the ring, i.e., $(G * h)(x) = \int_{-\pi}^{\pi} G(x-x') h(x') dx'$ for $2\pi$-periodic functions $G$ and $h$.
\par
We take $g(\omega)$ to be a Lorentzian distribution with half-width-at-half-maximum $D$, centered at $\omega=0$,
\begin{align}
	g(\omega) = \frac{1}{\pi}\frac{D}{\omega^2 + D^2},
\end{align}
for which the integral over $\omega$ in Eq.~\eqref{eq:W-a} can be performed by contour integration, giving $W(x, t) = \int G(x - x') a(x', -iD, t) dx'$.
Writing the first Fourier coefficient at $\omega = - i D$ as $\tilde{z}(x, t) = a(x, -iD, t)$, we obtain
\begin{align}
	\frac{\partial \tilde{z}}{\partial t} = -D \tilde{z} + \frac{i}{2}\left[ W e^{-i\beta} + \overline{W} e^{i\beta} \tilde{z}^2 \right],
	\qquad W = G * \tilde{z}.
\end{align}
The analysis is simplified by moving to a coordinate frame rotating at an as-yet-unknown rate $\Omega$.
Defining $z(x, t) = \tilde{z}(x, t) e^{i\Omega t}$, we find that $z$ satisfies
\begin{align}
	\frac{\partial z}{\partial t} = (i\Omega - D) z + \frac{i}{2}\left[ W e^{-i\beta} + \overline{W} e^{i\beta} z^2 \right] =: \mathcal{N}[z],
	\label{eq:oa}
\end{align}
where $\mathcal{N}[z]$ represents the dynamics of $z$. We will analyze this nonlocal field equation in what follows.
\par

Note that the quantity $r(x, t) = |z(x, t)| = |\tilde{z}(x, t)|$ characterizes the local coherence of the oscillator at $x$; $r = 1$ corresponds to perfectly locked oscillators and $0 \leq r < 1$ to partially coherent oscillators.
As shown in Laing~\cite{laing2009chimera}, the phase PDF $f$ at position $x$ is given by the Poisson kernel
\begin{align}
	f(x, \omega, \theta, t) = \frac{1 - r^2}{ 2 \pi [ 1 - 2 r \cos( \theta - \Theta ) + r^2 ] }
\end{align}
when $0 \leq r < 1$ and $f = \delta(\theta - \Theta)$ when $r=1$.

\subsection{The stationary chimera}

A stationary chimera state corresponds to a stationary solution $\hat{z}$ of Eq.~\eqref{eq:oa}, satisfying
\begin{align}
	\mathcal{N}[\hat{z}]
	=
	(i\Omega - D) \hat{z} + \frac{i}{2}\left[ \hat{W} e^{-i\beta} + \overline{\hat{W}} e^{i\beta} \hat{z}^2 \right] = 0,
	\quad \hat{W} = G * \hat{z} .
	\label{eq:oa-stat}
\end{align}
Note that if $\hat{z}(x)$ is a solution of Eq.~\eqref{eq:oa-stat} then so are $\hat{z}(x - X)$ for any $X$ and $e^{i\varphi}\hat{z}(x)$ for any $\varphi$.
These invariances are inherited from the statistical invariance of Eq.~\eqref{eq:model} under spatial translation on the ring and under global phase shifts $\theta \mapsto \theta + \varphi$ of all oscillators.
These symmetries are central to what follows.
Equation~\eqref{eq:oa-stat} is quadratic in $\hat{z}(x)$ at each point $x$.
Solving this and choosing the root with $|\hat{z}| \le 1$, we obtain
\begin{align}
	\hat{z}(x) = \frac{D - i\Omega - \sqrt{(D - i\Omega)^2 + |\hat{W}(x)|^2}}{i \overline{\hat{W}}(x) e^{i\beta}} .
	\label{eq:root}
\end{align}
\par

For the cosine kernel, Eq.~\eqref{eq:kernel}, the convolution with an even function $h(x) = h(-x)$ can be written in the form $(G * h)(x) = C_0 + C_1\cos x$, since
\begin{align}
  & (G * h)(x) = \frac{1}{2\pi}\int_{-\pi}^{\pi} (1 + A\cos(x - x')) h(x') dx' \cr
	 & = \frac{1}{2\pi}\int_{-\pi}^{\pi} h(x') dx' + \cos x\frac{A}{2\pi}\int_{-\pi}^{\pi} h(x')\cos x' dx' \cr
  & \qquad + \sin x\frac{A}{2\pi}\int_{-\pi}^{\pi} h(x')\sin x' dx'
\end{align}
and the last integral vanishes for even $h$.
Thus, if the chimera core is centered at $x=0$, $\hat{z}$ is an even function of $x$ and $\hat{W} = G * \hat{z}$ has a form
\begin{align}
	\hat{W}(x) = w_0 + w_1 \cos x,
	\label{eq:closure0}
\end{align}
where
\begin{align}
	w_0 = \frac{1}{2\pi} \int_{-\pi}^{\pi} \hat{z}(x) dx, \qquad
	w_1 = \frac{A}{2\pi} \int_{-\pi}^{\pi} \hat{z}(x) \cos x  dx .
	\label{eq:closure}
\end{align}
Substituting Eq.~\eqref{eq:root} into Eqs.~\eqref{eq:closure0} and~\eqref{eq:closure} gives two complex equations for the unknowns $w_0$, $w_1$, and $\Omega$.
Furthermore, fixing the argument of $w_0$ by requiring $\Im w_0 = 0$ provides one additional equation.
We can solve these equations for $w_0$, $w_1$, and $\Omega$ using a Newton-type method.
This self-consistency calculation was developed in Refs.~\cite{kuramoto2002coexistence, abrams2004chimera, laing2009chimera, mircheski2025spatial}.
In this study, we solve Eq.~\eqref{eq:oa-stat} directly rather than integrating Eq.~\eqref{eq:oa} forward in time, because the stationary state is attained only for the correct value of $\Omega$.
\par
Throughout this study, we set $A = 0.995$, $\beta = 0.16$ and $D = 0.004$, and discretize the spatial domain into $N = 1024$ evenly spaced grid points in the numerical analysis of the nonlocal field equation.
For these parameters, we numerically obtain $\Omega = -0.7276$.
The order parameter profile $r(x) = |\hat{z}(x)|$ of the self-consistent solution $\hat{z}$ is shown in the top panel of Fig.~\ref{fig:adjoint}.
Values of $r$ close to $1$ correspond to synchronized oscillator domains on the ring, whereas the region of lower $r$ values corresponds to incoherent oscillator domains.
We denote the center of this incoherent domain (chimera core) by $X_0$ ($X_0 = \pi$ in Fig.~\ref{fig:adjoint}).
In what follows, to characterize the position of the chimera core of a general evolving solution $z(x, t)$, we use the following circular mean on the ring weighted by the incoherence $1 - |z(x, t)|$:
\begin{align}
	X_{c}(t) = \arg \int_{-\pi}^{\pi} \left[ 1 - |z(x, t)| \right] e^{ix}  dx .
\end{align}
\section{Adjoint-based reduction}
\subsection{Two-parameter family of neutrally stable chimera states}

In this section, we develop an adjoint-based reduced description of the chimera state under weak perturbations.
As stated above, Eq.~\eqref{eq:oa-stat} is invariant under two types of transformations	inherited from the original model, Eq.~\eqref{eq:model}:
\begin{align}
	x \to x - X,
	\quad
	\theta \to \theta + \varphi,
\end{align}
where $X$ represents spatial translation on the ring, and $\varphi$ denotes the uniform phase shift of all oscillators.
Consequently, the stationary chimera state $\hat z$ is not an isolated fixed point of the unperturbed field equation~\eqref{eq:oa}, $\dot z = \mathcal{N}[z]$, but a single representative of a two-parameter family of stationary solutions.
We denote this family of solutions as
\begin{align}
	\hat z(x|X,\varphi) = e^{i\varphi} \hat z_0(x - X), \qquad X,\varphi \in [-\pi,\pi),
	\label{eq:group-action}
\end{align}
where $\hat z_0(x) = \hat z(x|0, 0)$ is a representative solution of Eq.~\eqref{eq:oa-stat}, i.e., a stationary solution of Eq.~\eqref{eq:oa} with $X=0$ and $\varphi=0$ in an appropriate frame rotating at the frequency $\Omega$.
\par
This situation is analogous to a periodic solution $x(t)$ of an autonomous limit-cycle oscillator, where $x(t-\tau)$ is also a solution for any $\tau$, and consequently dynamical reduction can be performed by projecting the system state onto the neutral phase mode, where the response of the phase variable is characterized by the sensitivity function obeying the system's adjoint linearized equation~\cite{kuramoto1984chemical, nakao2014phase, nakao2016phase, winfree1980geometry, ermentrout1996type}.
Such reduction to neutral modes has been performed in many studies, for example, for coupled oscillators, reaction-diffusion systems, and fluid systems with continuous symmetries~\cite{kuramoto1984chemical,ei2002motion,nakao2014phase, kawamura2013collective, kawamura2015phase, biktasheva2009computation, biktasheva2003wave, nakao2021phase, kawamura2008collective, kawamura2010phase, kawamura2014collective, kawamura2011collective, nishiura2002far, ozawa2026phase, fujii2025optimal, weerasinghe2019predicting, ozawa2021feedback, lober2014controlling, kawamura2007noise, fujii2025optimal, weerasinghe2019predicting, ozawa2021feedback}.
In this study, we formulate a reduction theory for stationary chimera states in parallel with the classical phase reduction theory for limit cycles~\cite{kuramoto1984chemical, nakao2014phase}.
\par

\subsection{Linearization and tangent functions}

First, we linearize Eq.~\eqref{eq:oa} about a stationary solution $\hat z(x|X,\varphi)$ characterized by $X$ and $\varphi$.
Writing $z(x,t) = \hat z(x|X,\varphi) + \eta(x,t)$ for small $\eta(x,t)$, the local complex order parameter is represented as
\begin{align}
	W(x,t) & = \int G(x-x')\big[\hat z(x'|X,\varphi) + \eta(x',t)\big] dx' \notag \\
	       & = \hat W(x|X,\varphi) + (G\ast\eta)(x),
\end{align}
where $\hat W(x|X,\varphi) = (G\ast \hat z)(x)$, and, to the first order in $\eta$,
\begin{align}
	z(x,t)^2 = \hat z(x|X,\varphi)^2 + 2\hat z(x|X,\varphi)\eta(x,t) + O(\eta^2).
\end{align}
Substituting these into Eq.~\eqref{eq:oa} and keeping only the terms linear in $\eta$, we obtain
\begin{align}
	\frac{\partial \eta}{\partial t} = \mathcal N\left[ \hat z(x|X,\varphi)+\eta \right] - \mathcal N \left[ \hat z(x|X,\varphi) \right] =: \hat L_{X,\varphi}\eta,
	\label{eq:linear}
\end{align}
where we defined a linear operator
\begin{align}
	\hat L_{X,\varphi} \eta & = J(x)\eta + \frac{i}{2} e^{-i\beta} (G \ast \eta)
	+ \frac{i}{2} e^{i\beta} \hat z(x|X,\varphi)^2 (G \ast \bar\eta), \nonumber                                \\
	J(x)                    & = (i\Omega - D) + i e^{i\beta}\overline{\hat W(x|X,\varphi)}\hat z(x|X,\varphi).
	\label{eq:linop}
\end{align}
The first term of Eq.~\eqref{eq:linop} arises from the local dynamics at $x$, while the remaining two arise from the nonlocal coupling with the kernel $G$.
\par
We assume that $\hat L_{X,\varphi}$ has exactly two zero eigenvalues, reflecting the two types of continuous symmetries, and that every other eigenvalue has a strictly negative real part, ensuring nearby states are attracted to this family of solutions.
These two zero eigenvalues of $\hat L_{X,\varphi}$ follow directly from differentiating the stationarity condition itself.
Since $\mathcal N[\hat z(x|X,\varphi)] = 0$ for any $X$ and $\varphi$, differentiating this equation with respect to $X$ gives
\begin{align}
	0 & = \frac{\partial}{\partial X}\mathcal N[\hat z(x|X,\varphi)]                                                                                                                 \cr
	  & = \lim_{\epsilon\to+0}\frac1\epsilon\Big(\mathcal N[\hat z(x|X+\epsilon,\varphi)] -\mathcal N[\hat z(x|X,\varphi)]\Big)                                                      \cr
	  & = \lim_{\epsilon\to+0}\frac1\epsilon\Big(\mathcal N\big[\hat z(x|X,\varphi) +\epsilon \partial_X \hat z(x|X,\varphi)+O(\epsilon^2)\big] \cr 
    & \qquad \qquad - \mathcal N[\hat z(x|X,\varphi)]\Big) \cr
	  & = \hat L_{X,\varphi}\big(\partial_X \hat z(x|X,\varphi)\big).
\end{align}
Thus, $\partial_X \hat z(x|X,\varphi)$ is an eigenfunction of $\hat L_{X,\varphi}$ with zero eigenvalue.
Similarly, differentiating with respect to $\varphi$ gives $\hat L_{X,\varphi}\big(\partial_\varphi \hat z(x|X,\varphi)\big) = (\partial/ \partial \varphi) \mathcal N[\hat z(x|X, \varphi)] = 0$, i.e., $\partial_\varphi \hat z(x|X,\varphi)$ is the other zero eigenfunction of $\hat L_{X,\varphi}$.
Using Eq.~\eqref{eq:group-action}, we denote these zero eigenfunctions as
\begin{align}
	v(x|X,\varphi) & = \partial_X \hat z(x|X,\varphi) = -e^{i\varphi}\hat z_{0}'(x-X), \nonumber \\
	w(x|X,\varphi) & = \partial_\varphi \hat z(x|X,\varphi) = i e^{i \varphi} \hat z_{0}(x - X),
	\label{eq:invariants}
\end{align}
where $'$ denotes the derivative of a function, so that
\begin{align}
	\hat L_{X,\varphi}  v(x|X,\varphi) = 0, \qquad \hat L_{X,\varphi}  w(x|X,\varphi) = 0.
	\label{eq:zeroeigenvalueeqs}
\end{align}
Here, the zero eigenfunctions $v(x|X,\varphi)$ and $w(x|X,\varphi)$ of $\hat L_{X,\varphi}$ represent the two tangent vectors of the chimera state $\hat z(x|X,\varphi)$ specified by the position and phase parameters $(X, \varphi)$, and perturbing $\hat z(x|X,\varphi)$ along either direction yields no restoring force to the original state $(X, \varphi)$; the perturbations in these neutral directions do not vanish because of the neutral stability.
Note that these functions reduce to $v_0(x) := v(x|0,0) = -\hat z'_0(x)$ and $w_0(x) := w(x|0,0) = i\hat z_0(x)$ at $(X, \varphi) = (0, 0)$, and $v(x|X, \varphi) = e^{i \varphi} v_0(x-X)$ and $w(x|X,\varphi) = e^{i \varphi} w_0(x-X)$ hold.
\par
In numerical calculations, we thus need to consider only the reference parameter $(X, \varphi) = (0, 0)$, then the results for other parameter values can be obtained simply by translating the result along $X$ or $\varphi$.
Discretizing the ring using $N$ grid points and separating the real and imaginary parts as $\eta = \eta_R + i\eta_I$, we obtain a real $2N\times2N$ linear system for $(\eta_R, \eta_I) \in \mathbb R^{2N}$ from Eq.~\eqref{eq:linop}, whose eigenvalues and eigenvectors can be evaluated numerically.
Figure~\ref{fig:adjoint} shows the two zero eigenfunctions $v_0(x)$ and $w_0(x)$ for $(X, \varphi) = (0, 0)$ obtained by numerically solving Eq.~\eqref{eq:zeroeigenvalueeqs}.
It can also be confirmed numerically that only two of the eigenvalues vanish and every other eigenvalue has a strictly negative real part.

\subsection{Position and phase functionals}
\label{sec:phase-functionals}

Our aim in this section is to develop an adjoint-based reduction theory in a similar manner to the phase reduction theory for limit cycles~\cite{brown2004phase,nakao2014phase}.
To this end, we first introduce two functionals that give the position and phase of a given chimera state.
Since the system has only two zero eigenvalues and all other eigenvalues have negative real parts, every initial state $z(x)$ sufficiently close to the family of stationary solutions is attracted to one of the stationary solutions $\hat z(x|X, \varphi)$ whose parameters $(X, \varphi)$ are determined by the initial state $z(x)$.
That is, every state $z(x)$ sufficiently close to the family $\hat z(x|X,\varphi)$ converges, under the unperturbed flow $\dot z = \mathcal{N}[z]$, to some specific member of the family as
\begin{align}
	\lim_{t\to\infty} \left\| z(x,t) - \hat z(x|\Psi\{z\}, \Phi\{z\}) \right\| = 0,
\end{align}
where $\| \cdot \|$ represents the $L_2$ norm, i.e., $\| f(x) \|^2 = \int_{-\pi}^{\pi} |f(x)|^2 dx$.
Here, we introduced two real functionals $\Psi\{z\},\Phi\{z\}$ of the field variable $z$, which we call the asymptotic position and phase.
These functionals are analogous to the asymptotic phase functional for limit-cycle solutions of partial differential equations, which can also be characterized as a principal Koopman eigenfunctional associated with a pure imaginary eigenvalue ~\cite{nakao2014phase,nakao2020spectral,nakao2021phase}.
\par
These functionals represent the parameter values $(X, \varphi)$ of the family of solutions that are asymptotically reached from the initial condition $z(x, 0) = z(x)$ ($-\pi \leq x \leq \pi$).
Thus, if we know the position and phase functionals $\Psi$ and $\Phi$, we can predict the position $X = \Psi\{z\}$ and phase $\varphi = \Phi\{z\}$ of the asymptotic chimera state from the initial field variable $z$.
Note that, for the stationary chimera state $\hat z(x|X,\varphi)$, $\Psi\{\hat z(x|X,\varphi)\}=X$ and $\Phi\{\hat z(x|X,\varphi)\}=\varphi$ hold by definition.
\par
When the perturbation is absent, the position $X(t)=\Psi\{z(x,t)\}$ and the phase $\varphi(t)=\Phi\{z(x,t)\}$ of the chimera state are constants of motion under the evolution of the system.
Indeed,
\begin{align}
	\dot X(t) & = \dot\Psi\{z(x,t)\} \notag                                                                                                                           \\
	          & = \lim_{\epsilon\to+0}\frac1\epsilon\Big[\Psi\{z(x,t+\epsilon)\} - \Psi\{z(x,t)\}\Big] \notag                                                         \\
	          & = \lim_{\epsilon\to+0}\frac1\epsilon\Big[\Psi\Big\{z(x,t)+\epsilon\frac{\partial z(x,t)}{\partial t}  +O(\epsilon^2)\Big\}-\Psi\{z(x,t)\}\Big] \notag \\
	          & = 2 \mbox{Re}\ \int \frac{\delta\Psi\{z\}}{\delta z(x,t)} \mathcal N[z(x,t)] dx 
            \cr
	          &= 2 \left\langle \overline{ \frac{\delta\Psi\{z\}}{\delta z(x,t)} }, \mathcal N[z(x,t)]\right\rangle = 0,
	\label{Eq:Psi0}
\end{align}
where the real inner product of two complex functions $z_1$ and $z_2$ is defined as
\begin{align}
	\langle z_1, z_2 \rangle = 	\mathrm{Re} \int_{-\pi}^{\pi} \overline{z_1(x)} z_2(x) dx,
	\label{eq:inner-product}
\end{align}
and $\delta \Psi\{z\} / \delta z(x, t)$ is the functional derivative of $\Psi\{z\}$ evaluated at $z = z(x, t)$.
Here, the functional derivative of a functional $F\{z\}$ of a complex field $z(x) = u(x) + i v(x)$ with real $u(x)$ and $v(x)$ with respect to $z(x, t)$ is defined as (see Appendix B)
\begin{align}
	\frac{\delta F\{z\}}{\delta z(x)} = \frac{1}{2} \left( \frac{\delta F\{z\}}{\delta u(x)} - i \frac{\delta F\{z\}}{\delta v(x)} \right).
\end{align}
Similarly, for the phase $\varphi$, we obtain
\begin{align}
	\dot\varphi(t) = \dot\Phi\{z(x,t)\}
	= 2 \left\langle \overline{ \frac{\delta\Phi\{z\}}{\delta z(x,t)} }, \mathcal N[z(x,t)]\right\rangle = 0.
	\label{Eq:Phi0}
\end{align}
Since $z(x,t)$ is an arbitrary state converging to the family of stationary chimera states, Eqs.~\eqref{Eq:Psi0} and \eqref{Eq:Phi0} should identically hold for $z(x, t)$, giving the exact identities
\begin{align}
	\left\langle \overline{ \frac{\delta \Psi\{z\}}{\delta z} }, \mathcal{N}[z] \right\rangle = 0,
	\qquad
	\left\langle \overline{ \frac{\delta \Phi\{z\}}{\delta z} }, \mathcal{N}[z] \right\rangle = 0,
	\label{eq:exact-conservation}
\end{align}
for every $z(x)$ in the basin of attraction of the solution family.

\subsection{Position and phase functionals and sensitivity functions}

It is generally difficult to obtain the functionals $\Psi$ and $\Phi$ explicitly, but we can linearly approximate them near the family of stationary chimera states.
For $z$ near $\hat z(x|X,\varphi)$, $\Psi\{z\}$ and $\Phi\{z\}$  can be expanded in functional Taylor series as
\begin{align}
	\Psi\{\hat z(x|X,\varphi) + \eta\}
	 & = \Psi\{\hat z(x|X,\varphi)\} + \cr 
  & 2 \mbox{Re} \int \left.\frac{\delta \Psi\{z\}}{\delta z(x)}\right|_{z = \hat z(x|X,\varphi)} \eta(x) dx + O(\eta^2) \cr
	 & = X + \langle { Q(x|X,\varphi) }, \eta(x) \rangle + O(\eta^2),
	\cr
	\Phi\{\hat z(x|X,\varphi) + \eta\}
	 & = \Phi\{\hat z(x|X,\varphi)\} + \cr 
  & 2 \mbox{Re} \int \left.\frac{\delta \Phi\{z\}}{\delta z(x)}\right|_{z = \hat z(x|X,\varphi)} \eta(x) dx + O(\eta^2) \cr
	 & = \varphi + \langle { R(x|X,\varphi) }, \eta(x) \rangle + O(\eta^2),
	\label{eq:taylor}
\end{align}
where we defined position and phase sensitivity functions as
\begin{align}
	Q(x|X,\varphi) & =
	2 \overline{\left.\frac{\delta \Psi\{z\}}{\delta z(x)} \right|}_{z(x) = \hat z(x|X,\varphi)},
	\cr
	R(x|X,\varphi) & =
	2 \overline{\left.\frac{\delta \Phi\{z\}}{\delta z(x)} \right|}_{z(x) = \hat z(x|X,\varphi)},
	\label{eq:QandR}
\end{align}
from the functional derivatives of $\Psi\{z\}$ and $\Phi\{z\}$ evaluated at $z(x) = \hat{z}(x | X, \varphi)$.
\par
As in phase reduction for limit cycles~\cite{brown2004phase, nakao2016phase, nakao2014phase}, these sensitivity functions can be calculated as the solutions to the adjoint equations without explicitly constructing the functionals $\Psi$ and $\Phi$.
Because $\Psi$ and $\Phi$ are conserved along the unperturbed evolution of the system, the position and phase functionals in Eq.~\eqref{eq:taylor} should stay constant along every solution $\eta(t, x)$ of the linearized equation~\eqref{eq:linear}.
For example, from the condition $d\Psi/dt = 0$, $Q(x|X,\varphi)$ should satisfy
\begin{align}
	0 & = \frac{d}{dt}\langle Q(x|X,\varphi), \eta(t)\rangle \notag          \\
	  & = \langle Q(x|X,\varphi), \hat L_{X,\varphi}\eta(t)\rangle \notag    \\
	  & = \langle \hat L_{X,\varphi}^\dagger Q(x|X,\varphi), \eta(t)\rangle,
\end{align}
and similarly $\langle \hat L_{X,\varphi}^\dagger R(x|X,\varphi), \eta(t)\rangle = 0$ should hold from the condition $d\Phi/dt = 0$.
Here, $\hat L_{X,\varphi}^\dagger$ is the adjoint operator of $\hat L_{X,\varphi}$ with respect to the real inner product, Eq.~\eqref{eq:inner-product}.
Since these equations should hold for every solution $\eta(t)$, the sensitivity functions $Q$ and $R$ should satisfy
\begin{align}
	\hat L_{X,\varphi}^\dagger Q(x|X,\varphi) = 0, \qquad \hat L_{X,\varphi}^\dagger R(x|X,\varphi) = 0,
	\label{eq:Ldagger}
\end{align}
namely, $Q$ and $R$ are the zero eigenfunctions of the adjoint operator $\hat L_{X,\varphi}^\dagger$.
We compute $\hat L_{X,\varphi}^\dagger$ explicitly in Sec.~\ref{sec:adjoint-calc} below.
\par
From Eq.~\eqref{eq:Ldagger}, $Q(x|X,\varphi), R(x|X,\varphi)$ can be fixed only up to a linear combination of the two.
Thus, we need to fix the remaining freedom from the bi-orthonormality conditions with the tangent vectors $v(x|X, \varphi)$ and $w(x|X, \varphi)$.
First, we differentiate the identity $\Psi\{\hat z(x|X,\varphi)\}=X$ by $X$ and $\varphi$.
Using the chain rule for functional differentiation, i.e.,
\begin{align}
	\frac{d}{dX} \Psi\{ z(x|X, \varphi) \}
	 & = \int 2 \frac{ \delta \Psi\{ z(x|X, \varphi) \} }{ \delta z(x|X, \varphi) }
	\frac{ \partial z(x|X, \varphi) }{ \partial X } dx
	\cr
	 & = \left\langle Q( x | X, \varphi ), \frac{ \partial z(x|X, \varphi) }{ \partial X } \right\rangle,
\end{align}
we obtain
\begin{align}
	\frac{d}{dX}\Psi\{\hat z(x|X,\varphi)\}       & = \big\langle Q(x|X,\varphi), \partial_X\hat z(x|X,\varphi)\big\rangle \notag \\
	                                              & = \langle Q(x|X,\varphi), v(x|X,\varphi)\rangle = 1,                          \\
	\frac{d}{d\varphi}\Psi\{\hat z(x|X,\varphi)\} & = \langle Q(x|X,\varphi), w(x|X,\varphi)\rangle = 0.
\end{align}
Similarly, differentiating $\Phi\{\hat z(x|X,\varphi)\}=\varphi$ by $X$ and $\varphi$ gives
\begin{align}
	\langle R(x|X,\varphi), v(x|X,\varphi) \rangle=0,
	\quad
	\langle R(x|X,\varphi), w(x|X,\varphi)\rangle=1.
\end{align}
Thus, we obtain the biorthogonality conditions
\begin{align}
	\langle Q(x|X,\varphi), v(x|X,\varphi)\rangle & = 1, \quad \langle Q(x|X,\varphi), w(x|X,\varphi)\rangle = 0, \\
	\langle R(x|X,\varphi), v(x|X,\varphi)\rangle & = 0, \quad \langle R(x|X,\varphi), w(x|X,\varphi)\rangle = 1,
	\label{eq:normalization}
\end{align}
which select $Q(x|X,\varphi)$ and $R(x|X,\varphi)$ uniquely from the two-dimensional function space of the zero eigenfunctions of $\hat L_{X,\varphi}^\dagger$.
\par
Note that, since $v(x|X, \varphi) = e^{i \varphi} v_0(x - X)$ and $w(x|X, \varphi) = e^{i \varphi} w_0(x-X)$, $Q(x|X, \varphi) = e^{i \varphi} Q_0(x - X)$ and $R(x|X, \varphi) = e^{i \varphi} R_0(x-X)$ hold, where $Q_0(x) = Q(x|0,0)$ and $R_0(x) = R(x|0,0)$.
Thus, we need to calculate the sensitivity functions only for $(X, \varphi)=(0, 0)$.
\par

\subsection{Calculation of the adjoint operator}
\label{sec:adjoint-calc}

We now explicitly compute the adjoint operator $\hat L_{X,\varphi}^\dagger$ of $\hat L_{X,\varphi}$ with respect to the real inner product, Eq.~\eqref{eq:inner-product}, satisfying $\langle L_{X,\varphi}^{\dag} a, \eta \rangle = \langle a, L_{X,\varphi} \eta \rangle$ for $2\pi$-periodic complex functions $a(x)$ and $\eta(x)$.
The calculation is identical for every stationary state $\hat z(x|X,\varphi)$ of the family, Eq.~\eqref{eq:linop}.
We drop the subscript of the linear operators and write $\hat z=\hat z(x|X,\varphi)$ throughout this section.
The linear operator $\hat L$ in Eq.~\eqref{eq:linop} is given by
\begin{align}
	\hat L\eta = J\eta + \tfrac i2 e^{-i\beta}(G*\eta) + \tfrac i2 e^{i\beta}\hat z^2(G*\overline\eta).
\end{align}
For the first term, we simply obtain $\langle a, J\eta\rangle = \langle \overline J a,\eta\rangle$.
For the second term, because $G$ is real and even, $G(x - x') = G(x' - x)$,
$
	\langle a, G*\eta \rangle
	= \mbox{Re}\ \int \overline{a(x)} (G*\eta)(x) dx
	= \mbox{Re}\ \int \int \overline{a(x)} G(x-x') \eta(x') dx' dx
	= \mbox{Re}\ \int \overline{ \int G(x'-x) a(x) dx } \eta(x') dx'
	= \mbox{Re}\ \int \overline{ (G*a)(x') } \eta(x') dx'
	= \langle G*a, \eta \rangle
$,
i.e., the convolution with $G$ is self-adjoint.
Together with the conjugation $\overline{\tfrac i2 e^{-i\beta}}=-\tfrac i2 e^{i\beta}$ of the complex coefficient, we obtain
$
	\left\langle a, \tfrac i2 e^{-i\beta}(G*\eta)\right\rangle = \left\langle -\tfrac i2 e^{i\beta}(G*a),\eta\right\rangle.
$
For the third term, writing $c=\tfrac i2e^{i\beta}\hat z^2$,
$
	\langle a, c \overline{(G*\eta)}\rangle
	= \operatorname{Re} \int \overline a c \overline{(G*\eta)} dx \notag
	= \operatorname{Re} \int\overline{a\overline c(G*\eta)} dx \notag
	= \operatorname{Re} \int a\overline c (G*\eta) dx \notag
	= \operatorname{Re} \int\big(G*(a\overline c)\big)\eta dx
	= \langle \overline{ G*(\overline a c) }, \eta \rangle
$
where we used $\operatorname{Re}\overline w=\operatorname{Re} w$ and that the convolution with $G$ is self-adjoint.
Since $G$ is real, $\overline{ G*(a\overline c) }={G*(\overline a c)}$, so we obtain
$
	\langle a, c \overline{(G*\eta)}\rangle =
	\big\langle G*(\overline a c), \eta\big\rangle = \left\langle \tfrac i2e^{i\beta} G* \big(\hat z^2\overline a\big), \eta\right\rangle.
$
Note that the coefficient $c = \tfrac i2e^{i\beta}\hat z^2$ is not conjugated as the above calculation shows.

Collecting these three contributions and restoring the $(X,\varphi)$ subscript, the adjoint operator of $\hat L_{X, \varphi}$ is given by
\begin{align}
	\hat L_{X,\varphi}^\dagger \xi = \overline J\xi - \frac i2 e^{i\beta}(G\ast\xi) + \frac i2 e^{i\beta} G\ast
	\left(\hat z^2\overline\xi\right).
	\label{eq:Ldagger-explicit}
\end{align}
The functions $Q(x|X,\varphi)$ and $R(x|X,\varphi)$ are the two zero eigenfunctions of this operator, fixed uniquely by the normalization and biorthogonality conditions as discussed in Sec.~\ref{sec:phase-functionals}.
In numerical computation, we consider the reference parameter $(X, \varphi) = (0,0)$, write $Q(x) = Q(x|0,0)$, $R(x) = R(x|0,0)$, $\hat L^\dagger=\hat L_{0,0}^\dagger$, and obtain $Q$ and $R$ numerically as the two zero eigenfunctions of the discretized operator.

\subsection{Reduction to position-phase dynamics}

We now derive the reduced equations for the position and phase of the chimera state under weak perturbations.
When the perturbation is absent, the position $X(t)=\Psi\{z(x,t)\}$ and the phase $\varphi(t)=\Phi\{z(x,t)\}$ remain constant under the evolution of the system.
Now let the chimera be weakly perturbed as
\begin{align}
	\frac{\partial z}{\partial t} = \mathcal{N}[z] + \varepsilon p(z, x, t),
	\label{eq:perturbed}
\end{align}
with $0 \leq \varepsilon \ll 1$, where $\varepsilon p(z,x,t)$ is a weak perturbation given to the nonlocal field equation that generally depends on the system state $z$, position $x$, and time $t$.
We assume that the perturbation is weak enough and does not destroy the chimera state, i.e., the perturbed state remains in the basin of attraction of the family of stationary solutions $\hat{z}(x | X, \varphi)$.

From Eqs.~\eqref{Eq:Psi0} and \eqref{Eq:Phi0}, if $\varepsilon\neq0$, the position $X(t)=\Psi\{z(x,t)\}$ and phase $\varphi(t)=\Phi\{z(x,t)\}$ obey
\begin{align}
	\dot X(t)      & = \left\langle 2 \overline{ \frac{\delta\Psi\{z\}}{\delta z(x,t)} }, \mathcal N[z(x,t)]+\varepsilon p(z(x,t),x,t)\right\rangle
	\cr
     & = \varepsilon \left\langle   2 \overline{ \frac{\delta \Psi\{z\}}{\delta z(x, t)} }, p(z(t),x,t)  \right\rangle,         \cr
%
	\dot\varphi(t) & = \left\langle 2 \overline{ \frac{\delta\Phi\{z\}}{\delta z(x,t)} }, \mathcal N[z(x,t)]+\varepsilon p(z(x,t),x,t)\right\rangle        
	\cr
	               & = \varepsilon  \left\langle 2 \overline{ \frac{\delta \Phi\{z\}}{\delta z(x,t)} }, p(z(t),x,t) \right\rangle,
	\label{eq:exact-reduced}
\end{align}
which are exact but not yet closed in $(X, \varphi)$ since the functional derivatives and the perturbation $p$ still depend on the full field $z(x, t)$.
\par
For sufficiently small perturbation $\varepsilon p(z, x, t)$, the perturbed state $z(x, t)$ remains in the neighborhood of the solution family $\hat z(x | X, \varphi)$, and we can approximately close Eq.~\eqref{eq:exact-reduced} in $(X, \varphi)$ at the lowest order in $\varepsilon$.
Writing $z(t) = \hat z(x|X(t),\varphi(t)) + \eta(t)$, we can assume $\eta(t)$ to be of $O(\varepsilon)$, because the non-zero eigenvalues of $\hat L_{X, \varphi}$ have negative real parts and deviations away from the solution family is exponentially suppressed.
Thus, we can drop $\eta(t)$ wherever it appears in the functional derivatives and in $p$, because they give rise only $O(\varepsilon^2)$ terms after expanding those quantities around $\hat z(x | X, \varphi)$.
\par

Combining Eq.~\eqref{eq:exact-reduced} with Eq.~\eqref{eq:QandR}, the position and phase of the perturbed chimera obey, at lowest order in $\varepsilon$,
\begin{align}
	\dot X(t)       & \approx \varepsilon\big\langle Q(x|X(t),\varphi(t)),  p(\hat z(x|X(t),\varphi(t)),x,t)\big\rangle, \cr
	\dot \varphi(t) & \approx \varepsilon\big\langle R(x|X(t),\varphi(t)),  p(\hat z(x|X(t),\varphi(t)),x,t)\big\rangle.
	\label{eq:reduced}
\end{align}
Here, all quantities are evaluated at $z(x) = \hat z(x|X(t),\varphi(t))$; therefore, these equations are closed in $(X, \varphi)$.
Thus, Eq.~\eqref{eq:reduced} reduces the infinite-dimensional field equation~\eqref{eq:perturbed} to two scalar ordinary differential equations for the position and phase of the chimera state under weak perturbations.\\
\par
\begin{figure}[htbp]
	\centering
	\includegraphics[width=\columnwidth]{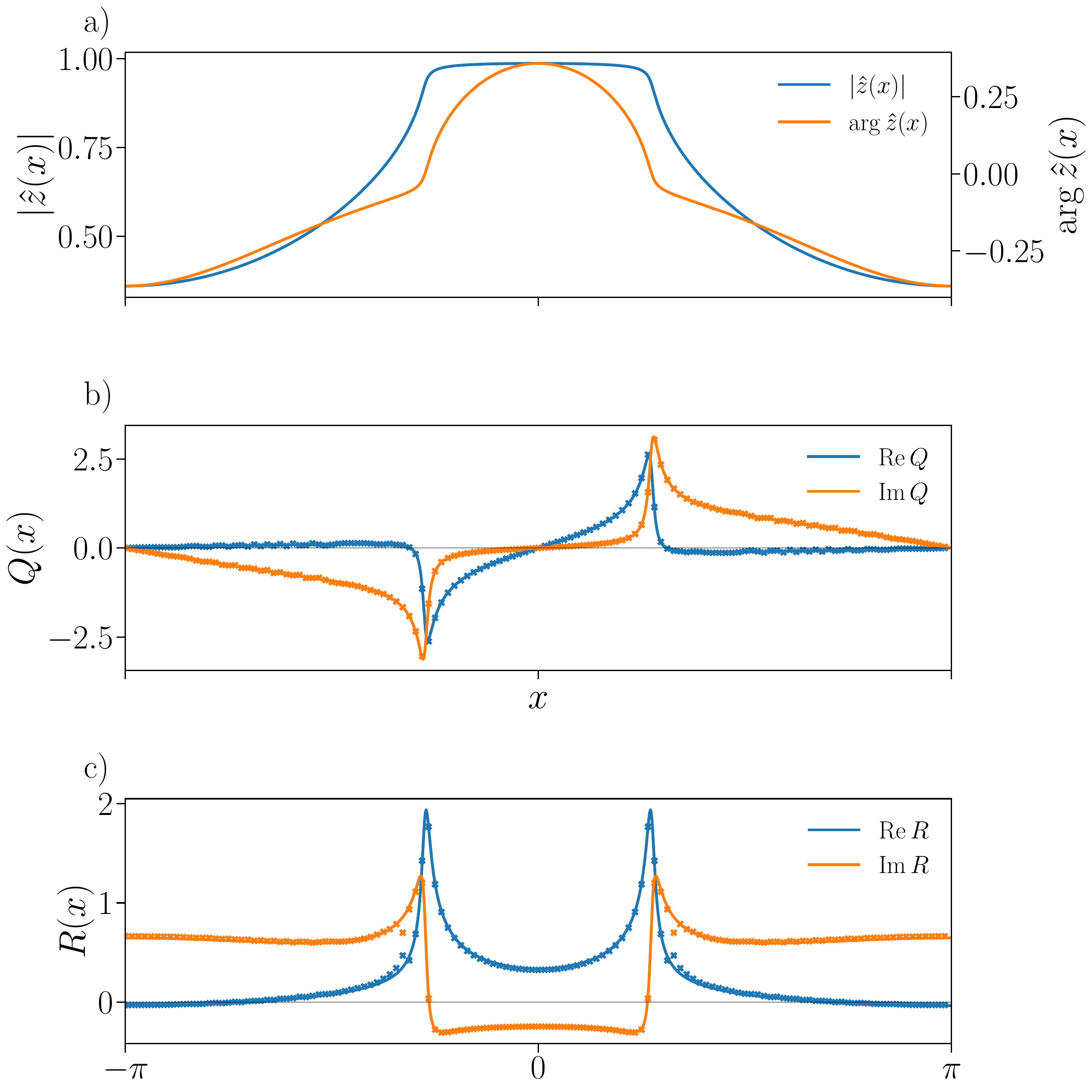}
	\caption{
		Stationary chimera, with the chimera core centered at $x = \pm\pi$, and its sensitivity functions.
		(a) Modulus $|\hat z(x)|$ (left axis) and argument $\arg \hat z(x)$ (right axis) of the stationary solution of Eq.~\eqref{eq:oa-stat}.
		(b) Position sensitivity function $Q(x)$ (real parts blue, imaginary parts orange).
		(c) Phase sensitivity function $R(x)$ (real parts blue, imaginary parts orange).
		Solid lines: the two zero eigenfunctions of the adjoint operator $\hat L^\dagger$.
		Crosses: direct measurement by impulsive kicks of the nonlocal field equation.
	}
	\label{fig:adjoint}
\end{figure}

\subsection{Sensitivity to microscopic perturbations}

In the previous subsection, $p(z, x, t)$ was a weak perturbation given to the nonlocal field equation.
Since our original model is a population of microscopic phase oscillators, Eq.~\eqref{eq:model}, rather than the nonlocal field equation, Eq.~\eqref{eq:oa}, we here consider the sensitivity of the system to perturbations given to the microscopic oscillators.
For phase oscillators, the only possibility to perturb them is via frequency modulation.
We thus add a driving field $\varepsilon u(x,t)$ directly to $\omega(x)$ in Eq.~\eqref{eq:model},
\begin{align}
	\dot\theta(x,t) & = \omega(x) + \varepsilon u(x,t) \cr
	                & +\int_{-\pi}^{\pi} G(x-x')\sin \big(\theta(x',t)-\theta(x,t)-\alpha\big) dx'.
\end{align}
Here, we assume that the perturbation $u(x, t)$ is smooth and affects all oscillators within each interval $[x, x+d x]$ in the same manner irrespective of their frequencies $\omega(x)$. 
\par
Since $u(x, t)$ uniformly acts on all oscillators in each interval $[x, x+d x]$, it changes the parameter $\omega$ in the phase velocity $v(x, \omega, \theta, t)$ in Eq.~\eqref{eq:v} from $\omega$ to $\omega + \epsilon u(x, t)$ at $x$.
It does not affect the contour integration that fixes $\omega = - i D$, and carries through every step from Eq.~\eqref{eq:dadt}
to Eq.~\eqref{eq:oa} as an extra term, giving
\begin{align}
	 &
	\frac{\partial z}{\partial t}
	= \mathcal N[z] - i\varepsilon u(x,t) z.
	\label{eq:oa-het}
\end{align}
Thus, by applying the frequency modulation $u(x, t)$ to the microscopic oscillators, the effective perturbation
\begin{align}
	p(z, x, t) = -i u(x,t) z,
\end{align}
to the nonlocal field equation can be realized in Eq.~\eqref{eq:reduced}.
\par
Substituting this $p$ into Eq.~\eqref{eq:reduced} gives
\begin{align}
	\dot X(t)
	 & = \langle
	Q(x|X, \varphi), -i \varepsilon u(x, t) \hat z(x|X, \varphi) \rangle
	\cr
	 & = \langle
	e^{i\varphi} Q_0(x-X), -i \varepsilon u(x, t) e^{i\varphi} \hat z_0(x - X) \rangle
	\cr
	 & = \varepsilon\int_{-\pi}^{\pi} u(x,t) S_Q(x-X(t)) dx,
	\label{eq:Spsi-def}
\end{align}
where we defined a position response kernel 
\begin{align}
	\dot X(t)
	 & = 
	 \varepsilon\ \langle Q(x|X, \varphi), -i  u(x, t) \hat z(x|X, \varphi) \rangle
	\cr
	 & = \varepsilon\ \mbox{Re}\ \int_{-\pi}^{\pi} -i  u(x,t) \overline{ Q(x|X, \varphi) } \hat{z}(x | X, \varphi) dx
\cr
	 & = \varepsilon\ \mbox{Re}\ \int_{-\pi}^{\pi} -i  u(x,t) \overline{ e^{i\varphi} Q_0(x-X) } e^{i \varphi} \hat{z}_0(x-X) dx
\cr
	 & = \varepsilon\ \mbox{Re}\ \int_{-\pi}^{\pi} -i \overline{ Q_0(x-X) } \hat{z}_0(x-X)  u(x,t) dx
\cr
	 & = \varepsilon\ \int_{-\pi}^{\pi} \mbox{Re}\ [ -i \overline{ Q_0(x-X) } \hat{z}_0(x-X) ] u(x,t) dx
\cr
	 & = \varepsilon\ \int_{-\pi}^{\pi} S_Q(x-X) u(x,t) dx
\end{align}
where $S_Q(x) =  \mbox{Re}\ [ -i \overline{ Q_0(x-X) } \hat{z}_0(x-X) = \mbox{Im}\ [ \overline{Q_0(x-X)} \hat{z}_0(x-X) ]$.
Unlike $Q$, the kernel $S_Q$ is real, as it must be to pair with a real driving field.
Similarly, for the phase $\varphi$, we obtain
\begin{align}
	 &
	\dot\varphi(t)=\varepsilon\int S_R(x-X(t)) u(x,t) dx,
	\cr
	 &
	S_R(x) = \mathrm{Im}[\overline{R_0(x - X)} \hat z_0(x - X)],
\end{align}
where the phase response kernel $S_R$ is also a real quantity.
\\
\par
\section{Numerical verification}

\subsection{Impulse response of the nonlocal field equation}

We have derived the reduced equations for the position and phase of the chimera state and defined the sensitivity functions $Q$ and $R$ to weak perturbations.
In this subsection, we numerically verify the sensitivity functions obtained from the adjoint equations by numerical simulations of the nonlocal field equation.
Suppose that the system rests at the reference chimera state $\hat z_{0}(x) = \hat z(x | 0,0)$ and receives a spatially localized impulsive kick at a single point $x_0$ at $t=0$,
\begin{align}
	z(x, 0) = \hat z(x) + \varepsilon c \delta(x - x_0), \qquad c \in \{1, i\},
	\label{eq:exact-kick}
\end{align}
where $c = 1$ and $c = i$ correspond to kicks in the real and imaginary parts of $z$ at $x_0$, respectively, and $\delta$ represents Dirac's delta, which is interpreted as a sufficiently narrow rectangular function of unit area in numerical calculations.
We assume that $\varepsilon$ is sufficiently small and the kicked state converges back to one of the stationary solutions $\hat z(x|X, \varphi)$ after relaxation.
\par
We perform direct numerical simulation of the nonlocal field equation~\eqref{eq:oa}, wait until a stationary chimera state is established, and perturb the system state as above.
Then we evolve the system and wait until the system state relaxes back to the stationary chimera state again.
We let $\delta X(x_0; \varepsilon c)$ and $\delta\varphi(x_0; \varepsilon c)$ denote the resulting position and phase shifts produced by the kick of Eq.~\eqref{eq:exact-kick} measured numerically.
The reduced pair of equations for the position and phase, \eqref{eq:reduced}, describe the linear response for small $\varepsilon$.
Therefore, the sensitivity functions $Q$ and $R$ can be represented as
\begin{align}
	\mathrm{Re} Q(x_0) = \lim_{\varepsilon \to 0} \frac{\delta X(x_0; \varepsilon)}{\varepsilon}, \qquad
	\mathrm{Im} Q(x_0) = \lim_{\varepsilon \to 0} \frac{\delta X(x_0; i\varepsilon)}{\varepsilon}, \\
	\mathrm{Re} R(x_0) = \lim_{\varepsilon \to 0} \frac{\delta \varphi(x_0; \varepsilon)}{\varepsilon}, \qquad
	\mathrm{Im} R(x_0) = \lim_{\varepsilon \to 0} \frac{\delta \varphi(x_0; i\varepsilon)}{\varepsilon}.
	\label{eq:direct-method}
\end{align}
Thus, by applying the real ($c=1$) and imaginary ($c=i$) kicks, the real and imaginary parts of each sensitivity function can be measured.
In numerical simulations, we use a finite but sufficiently small value of $\varepsilon$.
\par
Figure~\ref{fig:adjoint} compares the sensitivity functions $Q$ and $R$ measured numerically with those obtained from the adjoint equation.
Both sensitivity functions were measured at $128$ grid points of $x_0$, using kicks of the intensity $\varepsilon = 1 \times 10^{-2}$, and the resulting position and phase shifts were measured at $t = 400$, which is large enough for the system to relax back to the stationary chimera state.
The numerical results reproduce the predicted curves well over the whole ring.\\
\par
\subsection{Impulse response of the microscopic oscillators}

Next, we numerically verify the predicted response kernels $S_Q$ and $S_R$ of the chimera state to frequency perturbations given to the original microscopic oscillators, Eq.~\eqref{eq:model}.
We integrate a ring of $N=200\,000$ oscillators with $25$ different realizations of the natural frequencies drawn from the Lorentzian PDF $g(\omega)$ and generate a chimera state.
Then, as a localized perturbation $u(x, 0)$ at $t=0$, we split the ring into 256 blocks centered at $x_0$ and perturb the phases of all oscillators in $[x_0-w/2, x_0+w/2]$, a contiguous block of width $w = (2 / 250)\pi$, which contains $800$ oscillators, by $\Delta\theta=0.2$.
Such a block kick rotates the local complex order parameter around $x = x_0$, yielding an effective perturbation $-iu\hat z$ of Eq.~\eqref{eq:oa-het} with $u(x,0) \approx \delta(x - x_0)$.
The position response of the system is read out at $t=300$ after each kick.
Dividing the resulting displacement $\Delta X$ of the chimera state by the kick strength $\varepsilon=w\Delta\theta$, we obtain one sample of $S_Q$ at the position $x_0$ of the block.
The phase response kernel $S_R$ is also calculated in a similar manner by measuring the resulting phase shift $\Delta \varphi$.
\par
\begin{figure}[htbp]
	\centering
	\includegraphics[width=\columnwidth]{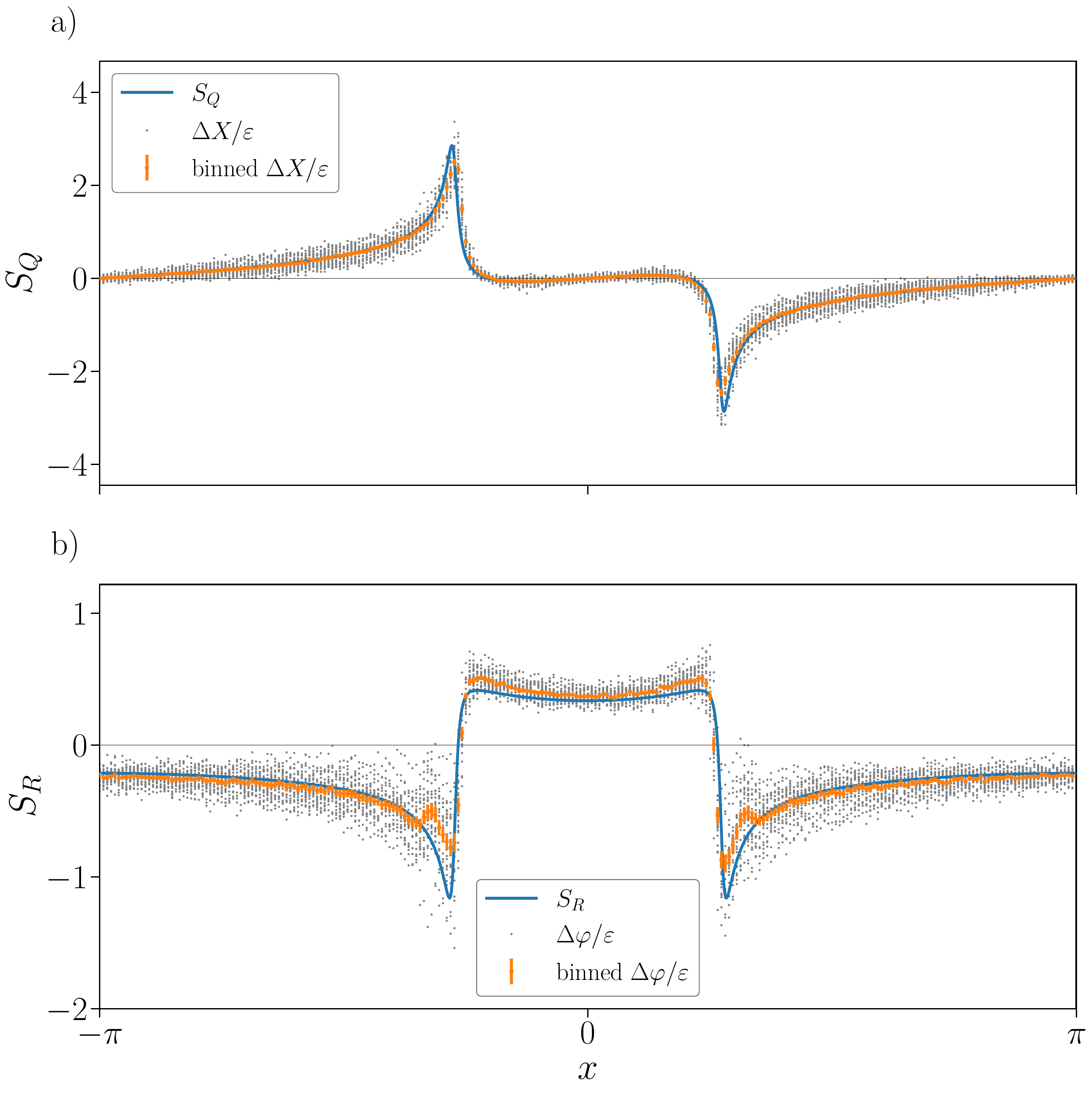}
	\caption{
		Position and phase response kernels $S_Q$ and $S_R$ of the chimera state of Eq.~\eqref{eq:model}, with $N = 200\,000$ oscillators, to microscopic perturbations measured by block phase kicks.
		(a) Position response kernel $S_Q$ of Eq.~\eqref{eq:Spsi-def}.
		(b) Phase response kernel $S_R$.
		Solid lines: kernels predicted from the adjoint equations.
		Grey dots: Position and phase response to individual kicks of $\Delta\theta = 0.2$ applied to one of $256$ blocks tiling the ring, in each of $25$ realizations of $\omega(x)$; the shift $\Delta X/\varepsilon$ or $\Delta\varphi/\varepsilon$ read out at $t = 300$ is plotted against the block's distance $x$ from the chimera core.
		Orange circles: the kicks binned by $x$, with the standard error of the mean as error bars.
	}
	\label{fig:micro}
\end{figure}

Figure~\ref{fig:micro} compares the position and phase responses averaged over the microscopic oscillators with the predicted position and phase kernels.
Each gray dot shows the response to a single kick; namely, the value $\Delta X/\varepsilon$ (panel a) or $\Delta\varphi/\varepsilon$ (panel b) of a single block in a single realization.
Their vertical spread reflects the finite-$N$ sample fluctuations.
The blue lines are the kernels $S_Q$ and $S_R$ calculated from the nonlocal field equation, with the core placed at $x=0$.
The orange symbols show the mean of each bin, plotted at the mean position of its kicks, with the standard error of the mean as error bars.
The averaged position and phase responses follow the predicted kernels reasonably well over the whole ring.
\par

\section{Position control of the chimera state}

\subsection{Position locking to a weak static field}

The reduced equations for $X$ and $\varphi$ in Eq.~\eqref{eq:reduced} predict how the position and phase of the chimera state respond to an applied weak field.
In this section, using the reduced equations, we consider a control problem for the position of the stationary chimera state.
\par
First, we take the perturbation field to be static, $p(z, x, t) = p(x)$, independent of the system state $z$ and time $t$.
Assuming that the phase parameter $\varphi$ is zero, the position $X$ of the chimera core obeys
\begin{align}
	\dot X = \varepsilon \Gamma(X), \qquad
	\Gamma(X) := \langle Q_{0}(x - X), p(x) \rangle,
	\label{eq:pos-coupling}
\end{align}
where $Q_0(x - X) = Q_0(x|X, 0)$.
The function $\Gamma(X)$ determining the dynamics of $X$ is $2\pi$-periodic and does not depend on $\varphi$.
If Eq.~\eqref{eq:pos-coupling} has a stable fixed point satisfying
\begin{align}
	\Gamma(X^\ast) = 0, \qquad \Gamma'(X^\ast) < 0,
\end{align}
the chimera core is pinned to the applied field at $X=X^*$.
Since $\Gamma$ depends linearly on the applied perturbation field $p$, we can easily design an appropriate functional form of $p$ to place $X^\ast$ anywhere on the ring.
\par
\begin{figure}[htbp]
	\centering
	\includegraphics[width=0.85\columnwidth]{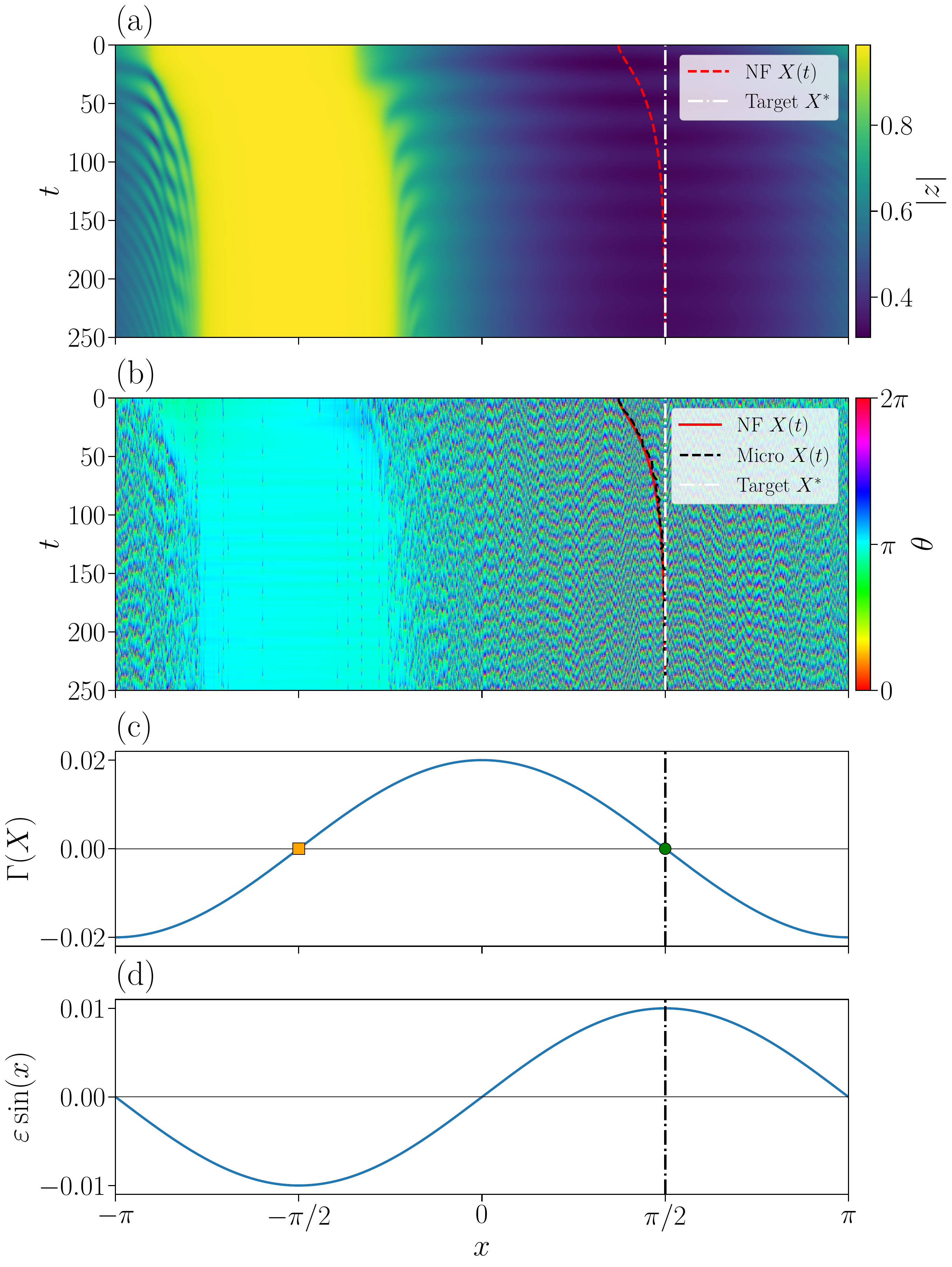}
	\caption{
		Locking of the chimera core by the sinusoidal heterogeneity.
		(a) Space-time plot of the coherence $|z(x,t)|$ from the nonlocal field equation, Eq.~\eqref{eq:oa-het}, with the core position $X(t)$ (red dashed) and the target $X^\ast$ (white dotted).
		(b) Phases $\theta(x,t)$ of a ring of $N = 100\,000$ oscillators, Eq.~\eqref{eq:model}, under the same forcing, with the measured core position (black dashed) and the core position from (a) (red).
		(c) Locking function $\Gamma(X)$ of Eq.~\eqref{eq:pos-coupling}, with its stable (circle) and unstable (square) zeros.
		(d) Applied heterogeneity $\varepsilon b(x)$.
	}
	\label{fig:heterogeneity}
\end{figure}

We perform direct numerical simulations of the original oscillator model, Eq.~\eqref{eq:model}.
For a static profile $u(x,t)=b(x)$ of the frequency modulation applied to the microscopic oscillators, Eq.~\eqref{eq:Spsi-def} gives $\dot X=\varepsilon\Gamma_b(X)$ with $\Gamma_b(X) = \langle S_Q(x-X), b(x) \rangle = \langle Q_0(x-X), -i b(x) \hat z_0(x-X) \rangle$.
As the simplest example, we take the sinusoidal profile, $b(x) = \sin x$, and set $\varepsilon = 0.01$, shown in Fig.~\ref{fig:heterogeneity}(b).
In this case, the system has a single stable fixed point at $X^\ast=\pi/2$ and a single unstable fixed point diametrically opposite it at $-\pi/2$ as shown in Fig.~\ref{fig:heterogeneity}(c).
Wherever the chimera core starts (except the unstable fixed point $X_0=-\pi/2$), it eventually locks to the applied frequency-modulation field at $X^\ast$.
Figures~\ref{fig:heterogeneity}(a) and (b) confirm this for a chimera core released at $X(0) = X^\ast - 0.4$, both in the nonlocal field equation, Eq.~\eqref{eq:oa-het} (Fig.~\ref{fig:heterogeneity}(a)), and in the direct simulation of $N = 100\,000$ oscillators (Fig.~\ref{fig:heterogeneity}(b)), showing good agreement between the results of the original oscillator model and the nonlocal field equation.
The red dashed line in Fig.~\ref{fig:heterogeneity}(a) and the solid red line in Fig.~\ref{fig:heterogeneity}(b) show the core position $X(t)$ of the nonlocal field equation; the black dashed line in Fig.~\ref{fig:heterogeneity}(b) shows the core position measured from the direct simulation; and the dash-dotted vertical line marks the target $X^\ast$; the two trajectories are in great agreement throughout the relaxation to $X^\ast$.

\subsection{Optimal input field for fast control}

We can further use the reduced equation to derive the optimal input field for fast control of the chimera core.
We find the optimal profile $u(x)$ of the frequency modulation that locks the chimera core at a chosen target $X^\ast$ satisfying $\Gamma_u(X^\ast)=0$ under a fixed control energy $\int u(x)^2 dx = E$, which realizes the fastest possible relocation of the chimera core to $X^\ast$ within linear approximation.
We follow a similar constrained-optimization approach used by Zlotnik \textit{et al.}~\cite{zlotnik2013optimal} for the optimal entrainment of a periodically forced oscillator.

The linear stability of the fixed point $X^\ast$ is determined by the slope $\Gamma_u'(X^\ast)$ of $\Gamma_u = \langle S_Q(x-X), u(x) \rangle$ at $X^\ast$.
We thus consider an optimization problem
\begin{align}
	\max_{u}\ -\Gamma_u'(X^\ast)
	\qquad
	\text{s.t.}\quad \Gamma_u(X^\ast)=0,\quad \int_{-\pi}^{\pi} u(x)^2 dx = E.
\end{align}
As in Ref.~\cite{zlotnik2013optimal, takata2021fast}, we introduce a Lagrange multiplier $\mu$ for the constraint, and consider the cost functional
\begin{align}
	C(u,\mu) = -\Gamma_u'(X^\ast) + \mu\Big(E - \int u^2 dx\Big).
\end{align}
Here, differentiating $\Gamma_u(X)$ gives
%
\begin{align}
	\Gamma_u'(X) = -\big\langle S_Q'(x-X), u(x)\big\rangle.
	\label{eq:gamma-prime}
\end{align}
The stationary condition $\delta C /\delta u = 0$ gives, using Eq.~\eqref{eq:gamma-prime},
\begin{align}
	S_Q'(x-X^\ast) - 2\mu u(x) = 0,
\end{align}
from which we derive
\begin{align}
	u(x) = \frac{1}{2\mu}\Big[S_Q'(x-X^\ast)\Big].
\end{align}
The energy constraint then fixes $\mu$,
\begin{align}
	\mu = \frac{||S_Q'|| }{2 \sqrt E},
\end{align}
where the positive root is chosen so that $-\Gamma_u'(X^\ast) > 0$.
Thus, the optimal actuation is derived as
\begin{align}
	u(x) = \varepsilon S_Q'(x-X^\ast), \qquad \varepsilon = \frac{\sqrt E}{\|S_Q'\|},
	\label{eq:uopt}
\end{align}
which attains the maximal steepness around the locking point $X^\ast$ achievable with a fixed energy $E$.
\par
We now confirm that $X^\ast$ is indeed a fixed point.
Substituting Eq.~\eqref{eq:uopt} into $\Gamma_u(X)$ and shifting the integration variable, we obtain
\begin{align}
	\big\langle S_Q(x-X^*), S_Q(x-X^*)'\big\rangle
	= \tfrac12 \int_{-\pi}^{\pi} \frac{d}{dx} \{ S_Q(x)^2 \} dx = 0.
\end{align}
This means that the external field $u(x) \propto S_Q'(x-X^\ast)$ already satisfies the locking condition on its own.
\par
\begin{figure}[htbp]
	\centering
	\includegraphics[width=0.85\columnwidth]{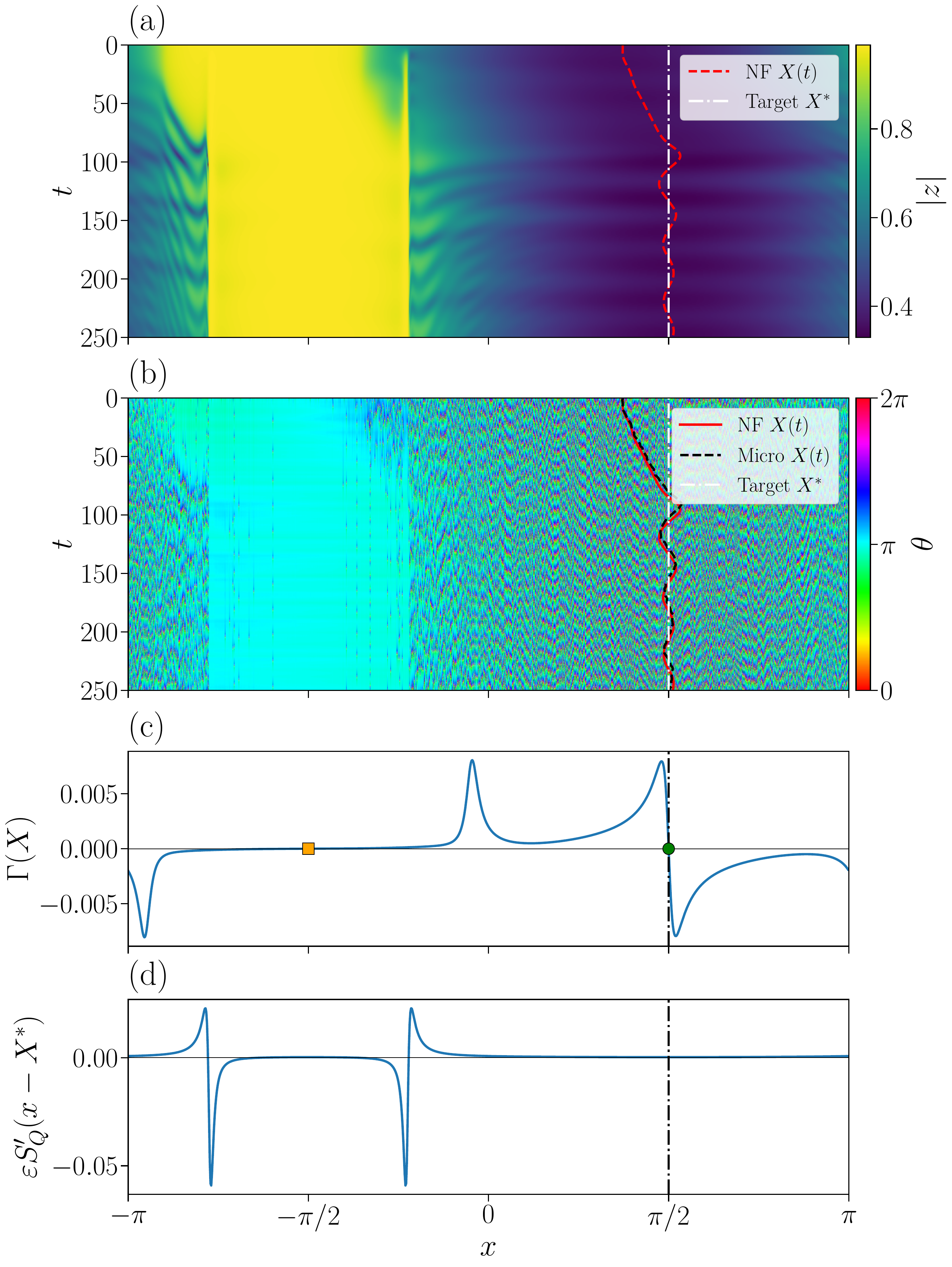}
	\caption{
		Locking of the chimera core by the optimal input field $u(x) = \varepsilon S_Q'(x - X^\ast)$.
		(a) Space-time plot of the coherence $|z(x,t)|$ from Eq.~\eqref{eq:oa-het}, with the core position $X(t)$ (red dashed) and the target $X^\ast$ (white dotted).
		(b) Phases $\theta(x,t)$ of a ring of $N = 100\,000$ oscillators under the same forcing, with the measured core position (black dashed) and the core position from (a) (red).
		(c) Locking function $\Gamma(X)$, with its stable (circle) and unstable (square) zeros.
		(d) Optimal input field $\varepsilon S_Q'(x - X^\ast)$.
	}
	\label{fig:pinning}
\end{figure}

Figure~\ref{fig:pinning} shows the spatial locking of the chimera core by the optimal $u(x)$, starting from the same initial state and locking to the same target as Fig.~\ref{fig:heterogeneity} with the sinusoidal $b(x)$.
Both inputs have the same $L^2$ norm on the ring, i.e., the same control energy $E = \pi\times10^{-4}$.
In both cases, the system smoothly settles onto $X^\ast$.
We can confirm that, in the linear regime close to $X^\ast$, the optimal field relocates the chimera core faster.
The lines in Fig.~\ref{fig:pinning} have the same meaning as in Fig.~\ref{fig:heterogeneity}, and the core trajectory of the oscillator network is in agreement with the nonlocal field equation.

\section{Conclusions}

We have derived an adjoint-based reduced description for the position and phase of a stationary chimera state in a nonlocally coupled ring of Kuramoto-Sakaguchi oscillators under weak perturbation.
Starting from the nonlocal field equation derived via the Ott-Antonsen ansatz, we treated the stationary chimera as a two-parameter family of stationary solutions stemming from the system's two continuous symmetries with respect to the spatial-translation on a ring and global phase-shift of the oscillators.
Linearizing about this family of solutions and constructing the adjoint of the resulting operator, we obtained a position-sensitivity function and a phase-sensitivity function as its two zero eigenfunctions, fixed uniquely by biorthogonality against the tangent functions of the solution family.
These two functions characterize the linear response of the infinite-dimensional field variable of the chimera state to a weak perturbation, yielding a pair of ordinary differential equations for the position and the phase of the chimera core.
We formulated this theory in parallel with the phase reduction of a limit cycle, which has a similar symmetry-generated neutral mode.
\par
We verified the reduced equations by numerical simulations of the nonlocal field equation, and also microscopically by direct numerical simulation of the original oscillator model.
We then used the reduced equations for control of the chimera state.
We showed that a weak static field can lock the position of the chimera at the stable zero of the position-coupling function derived from the reduced equation, and also derived the optimal field that maximizes the linear stability, yielding fast locking of the chimera to the applied field.
\par
The present reduction framework relies on the assumptions that the chimera state is an attracting stationary state, with exactly two neutral directions, and that the perturbation is weak enough that the system state stays close to the family of stationary solutions.
Within these limits, the present framework is not tied to the particular kernel or parameter values used here.
Several future extensions follow naturally, the corresponding reduction for breathing and traveling chimeras, where the position is no longer neutral.

\section*{Acknowledgements}
We acknowledge JSPS KAKENHI Nos. 25H01468, 25K03081, and 22H00516 for financial support.

\section*{Data Availability Statement}
The data that support the findings of this study are available within the article.

\section*{Artificial Intelligence (AI) declaration}
During the preparation of this work, AI has been used to rephrase a few sentences into more formal English with the goal of improving readability.
The authors take full responsibility for the content of the publication.
\appendix

\section{Expression of the local complex order parameter}

Here, we formally derive the expression for the local complex order parameter $W(x, t)$.
We discretize the interval $[-\pi, \pi]$ using $N$ grid points, $x_n = n \Delta x - \pi$ ($n=0, ..., N-1$) where $\Delta x = 2 \pi / N$.
Then, assuming $G(x-x')$ to be a smooth function, $W(x, t)$ can be approximated as
\begin{align}
    & W(x, t)
      =  \int_{-\pi}^{\pi} G(x-x') e^{-i \theta(x')} dx'
    \cr
     & =  \lim_{N \to \infty} \sum_{n=0}^{N-1} \int_{x_n}^{x_n + \Delta x} G(x-x') e^{-i \theta(x')} dx'
    \cr
     & =  \lim_{N \to \infty} \sum_{n=0}^{N-1} \int_{0}^{\Delta x} G(x -x_n -h) e^{-i \theta(x_n+h)} dh
    \cr
     & =  \lim_{N \to \infty} \sum_{n=0}^{N-1} \int_{0}^{\Delta x} [ G(x -x_n) + O(\Delta x) ] e^{-i \theta(x_n+h)} dh
    \cr
     & =  \lim_{N \to \infty} \left( \sum_{n=0}^{N-1} G(x-x_n) \int_{0}^{\Delta x} e^{-i \theta(x_n + h)} dh + O(N \Delta x^2) \right),
\end{align}
where the $O(N \Delta x^2) = O(1/N)$ term vanishes as $N \to \infty$.
\begin{align}
    \int_0^{\Delta x} e^{-i \theta(x_n + h)} dh
    =
    \lim_{M \to \infty}
    \frac{1}{M}
    \sum_{k=0}^{M-1}
    e^{-i \theta_k}
    \Delta x,
\end{align}
where $\theta_k = \theta(x_n + \Delta x / M)$.
Assuming that infinitely many oscillators exist in the interval $[x_n, x_n+\Delta x]$ with phase values distributed according to $f(x_n, \omega, \theta, t)$, we represent this quantity as an average over $\omega$ and $\theta$ as
\begin{align}
    \lim_{M \to \infty}
    \frac{1}{M}
    \sum_{k=0}^{M-1}
    e^{-i \theta_k}
    \Delta x
    =
    \int_{-\pi}^{\pi} \int_{-\infty}^{\infty} f(x_n, \omega', \theta', t) e^{-i \theta'} d\theta' d\omega' \Delta x.
\end{align}
Thus, we obtain
\begin{align}
    W(x, t)
     & =  \lim_{N \to \infty} \left( \sum_{n=0}^{N-1} G(x-x_n) \int_{-\pi}^{\pi} \int_{-\infty}^{\infty} f(x_n, \omega', \theta', t) e^{-i \theta} d\theta' d\omega' \Delta x \right)
    \cr
     & =
    \int_{-\pi}^{\pi} G(x-x') \int_{-\pi}^{\pi} \int_{-\infty}^{\infty} f(x_n, \omega', \theta', t) e^{-i \theta} d\theta' d\omega'.
\end{align}

\section{Functional derivatives and real inner products}

Here, we briefly summarize the notions for functional derivatives and real inner products of real-valued functionals of complex functions used in this study.
The functional derivative of a real functional $F\{y\}$ of a real function $y(x)$ is defined as usual via the G\^ateaux derivative:
\begin{align}
\lim_{\epsilon\to+0}
\frac{F\{ y+\epsilon \eta\} - F\{ y \}}{\epsilon}
=
\int \frac{\delta F\{y\}}{\delta y(x)} \eta(x) dx.
\end{align}
We consider a real-valued functional $G\{ z \}$ of a complex function $z(x) = u(x) + i v(x)$, where $u(x)$ and $v(x)$ are two real functions, which is equal to a functional $G\{ u, v \}$ of $u(x)$ and $v(x)$, $G\{ z \} = G\{ u, v \}$.
The functional derivative of $G\{ u, v \}$ is defined via
\begin{align}
&\lim_{\epsilon\to+0}
\frac{G\{u+\epsilon \eta_u, v+\epsilon \eta_v\} - G\{u, v\}}{\epsilon}
= \\ \cr
&\int \left( \frac{\delta G\{u, v\}}{\delta u(x)} \eta_u(x) + \frac{\delta G\{u, v\}}{\delta v(x)} \eta_v(x) \right) dx.
\end{align}
Following the Wirtinger calculus, we introduce the functional derivative of a real functional $G\{ z \} = G\{ u, v \}$ of a complex function $z(x)$ as
\begin{align}
\frac{\delta G(z)}{\delta z(x)} = 
\frac{1}{2} \left( \frac{\delta G\{u, v\}}{\delta u(x)} - i \frac{\delta G\{u, v\}}{\delta v(x)} \right).
\end{align}
Then, writing $\zeta(x) = \zeta_u(x) + i \zeta_v(x)$ where $\zeta_u(x)$ and $\zeta_v(x)$ are real functions, we have
\begin{align}
&\int \frac{\delta G\{z\}}{\delta z(x)} \zeta(x) dx = \cr 
& \int \frac{1}{2} \left( \frac{\delta G\{u, v\}}{\delta u(x)} - \frac{\delta G\{u, v\}}{\delta v(x)} \right) ( \zeta_u(x) + i \zeta_v(x) ) dx =  \cr
& \frac{1}{2}  \int  \Big\{ \left( \frac{\delta G\{u, v\}}{\delta u(x)} \zeta_u(x) + \frac{\delta G\{u, v\}}{\delta v(x)} \zeta_v(x) \right) \cr
&\quad + i \left( \frac{\delta G\{u, v\}}{\delta u(x)} \zeta_v(x) - \frac{\delta G\{u, v\}}{\delta v(x)} \zeta_u(x)  \right) \Big\} dx.
\end{align}
Therefore, 
\begin{align}
&\lim_{\epsilon\to+0}
\frac{G\{ z + \epsilon \zeta \} - G\{ z \}}{\epsilon}
= \nonumber \\ 
&\lim_{\epsilon\to+0}
\frac{G\{ u+\epsilon \zeta_u, v+\epsilon \zeta_v \} - G\{ u, v \}}{\epsilon}
= \nonumber \\ 
& 2 \mbox{Re}\
\int \frac{\delta G\{z\}}{\delta z(x)} \zeta(x) dx
\end{align}
holds, which is used in the functional Taylor expansion in Eq.~\eqref{Eq:Psi0} and other equations.

\bibliography{refs}

\end{document}